Review

# Machine learning for modeling the progression of Alzheimer disease dementia using clinical data: a systematic literature review

Sayantan Kumar[1], Inez Oh[1], Suzanne Schindler[2], Albert M. Lai [1],
Philip R.O. Payne [1], and Aditi Gupta[1]

[1]Institute for Informatics, Washington University School of Medicine, St. Louis, Missouri, USA and [2]Department of Neurology, Washington University School of Medicine, St. Louis, Missouri, USA

Corresponding Author: Aditi Gupta, PhD, Institute for Informatics, Washington University School of Medicine, Campus Box 8102, 660 S Euclid Ave, St. Louis, MO 63110, USA; agupta24@wustl.edu

## ABSTRACT

**Objective:** Alzheimer disease (AD) is the most common cause of dementia, a syndrome characterized by cognitive impairment severe enough to interfere with activities of daily life. We aimed to conduct a systematic literature review (SLR) of studies that applied machine learning (ML) methods to clinical data derived from electronic health records in order to model risk for progression of AD dementia.
**Materials and Methods:** We searched for articles published between January 1, 2010, and May 31, 2020, in PubMed, Scopus, ScienceDirect, IEEE Explore Digital Library, Association for Computing Machinery Digital Library, and arXiv. We used predefined criteria to select relevant articles and summarized them according to key components of ML analysis such as data characteristics, computational algorithms, and research focus.
**Results:** There has been a considerable rise over the past 5 years in the number of research papers using ML-based analysis for AD dementia modeling. We reviewed 64 relevant articles in our SLR. The results suggest that majority of existing research has focused on predicting progression of AD dementia using publicly available datasets containing both neuroimaging and clinical data (neurobehavioral status exam scores, patient demographics, neuroimaging data, and laboratory test values).
**Discussion:** Identifying individuals at risk for progression of AD dementia could potentially help to personalize disease management to plan future care. Clinical data consisting of both structured data tables and clinical notes can be effectively used in ML-based approaches to model risk for AD dementia progression. Data sharing and reproducibility of results can enhance the impact, adaptation, and generalizability of this research.

## INTRODUCTION

Alzheimer disease (AD) is the most common cause of dementia, which is a syndrome characterized by impairment of memory and/or thinking severe enough to interfere with activities of daily life.[1,2] AD dementia affects millions of people worldwide and is currently the sixth leading cause of death in the United States.[3,4] AD-related brain pathology, which includes the accumulation and deposition of amyloid-β peptide and tau protein, begins almost 10–20 years before the onset of dementia symptoms.[5] Therefore, many individuals with early AD brain pathology are cognitively normal but at higher risk for developing dementia in the future.[6] AD dementia is progressive and incurable, and, at advanced stages, patients suffer potentially

**LAY SUMMARY**
Alzheimer disease (AD) is the most common cause of dementia, which is a syndrome of impaired memory and/or thinking that interferes with activities of daily life. Many studies of AD dementia utilize information from expensive and invasive procedures, such as brain imaging or spinal taps, to estimate risk for developing AD dementia or rapid cognitive decline. However, widely available electronic health records (EHRs) systems contain a wealth of healthcare data describing a patient's medical history and clinical presentation (eg, demographics, vital signs, medications, laboratory data, current medical conditions) that could be leveraged as a low-cost, noninvasive alternative to study the progression of AD dementia. In recent years, machine learning (ML) has become a useful tool in identifying hidden patterns within large-scale healthcare data, such as the aforementioned EHR-derived data types, leading to increased efficiency and improved healthcare. We aim to perform a systematic literature review of studies applying ML to clinical and EHR data to identify factors that predict risk for progression of AD dementia. We summarize the reviewed articles according to key components of ML analysis such as data characteristics, computational algorithms, and research focus. Finally, we identify gaps in the literature and potential opportunities for future research.

fatal complications such as dehydration, malnutrition, or infection.[7] Identifying individuals with early AD brain pathological changes could lead to therapeutic interventions to delay the disease progression over time and could be helpful for tailoring disease management and planning future care.

Clinical data are defined as "information ranging from determinants of health and measures of health and health status to documentation of care delivery … captured for a variety of purposes and stored in numerous databases across the healthcare system."[8] Nonimaging clinical data extracted from EHRs are some of the most accessible and widely used clinical datasets. They are an integral part of contemporary healthcare delivery, enabling quick access to accurate, up-to-date, and complete patient information, and assisting in accurate diagnosis and coordinated, efficient care.[9] EHR data collected from individuals at risk for AD dementia can include laboratory test results, vital signs, medications, and other treatments administered, as well as comorbidities.[9,10] In some cases, patients may also undergo specific testing for markers linked to AD brain pathology using expensive and/or invasive procedures such as neuroimaging scans (magnetic resonance imaging [MRI] and position emission tomography [PET]) and cerebrospinal fluid (CSF) collection for biomarker testing.[11–15] The results of these tests may also be present in the EHR. Research has shown that such longitudinal clinical EHR data (ie, data collected from multiple time points) can be utilized for monitoring the time-course of AD dementia progression.[16]

The widespread use and availability of medical devices over the years have provided an overwhelming volume of clinical EHR data, which could potentially augment the traditional tools of dementia experts.[17] The unmet needs for dementia expertise, coupled with the relevant massive datasets, have encouraged researchers to examine the utility of artificial intelligence (AI), which is gaining high visibility in the realm of healthcare innovation.[18] Machine learning (ML), a branch of AI, can model the relationship between the input quantities and clinical outcomes, discover hidden patterns within large-scale data, and make inferences or decisions that help in more accurate clinical decision-making.[19] However, computational hypotheses generated by ML models still need to be validated by subject matter experts in order to ensure adequate precision for clinical decision-making purposes.[20] In our review, we include studies using ML for the purpose of predictive modeling (such as decision trees, support vector machines [SVM], k-means clustering) and exclude studies using statistical methods for cohort summarization and hypothesis testing (such as odds ratio, Chi-square distribution, Kruskal–Wallis test, Kappa-Cohen test). For studies using linear and logistic regression only those studies were included which utilized these methods for predictive modeling or classification analysis.

The current literature generally neglects secondary use of nonimaging clinical data, including routinely collected EHR-derived data, as a rich, low-cost, and noninvasive source of information for identifying potential risk factors for AD dementia, and instead focuses on the use of costlier and/or invasive imaging and diagnostic testing data for ML-based analysis. However, we believe that further study of EHR-derived data could lead to more efficient, cost-effective, timely, and personalized disease management for individuals with AD. With this motivation and a goal of identifying the knowledge gaps and potential opportunities for the use of EHR-derived data in conjunction with ML frameworks, our goal was to conduct a systematic literature review (SLR) on the state-of-the-art of ML as applied EHR-derived data for the purposes of modeling and understanding AD dementia progression.

## METHODS

As noted above, the motivation behind conducting an SLR is to summarize existing findings related to a chosen research topic to identify gaps in literature and thus create a ground for future research work. As stated by Martí-Juan et al,[19] "Performing an SLR comprises the following steps: (1) identify the need for performing the SLR; (2) formulate research questions; (3) execute a comprehensive search and selection of primary studies; (4) assess the quality and extract data from the studies; (5) interpret the results; and (6) report the SLR." In this SLR, our main research question is: *How are machine learning algorithms being applied by researchers for studying progression of AD dementia using clinical EHR data?* This main question can be subdivided further into the following 3 research questions:

1. What type of ML methods have been used for detecting the onset of AD dementia and for predicting the trajectory of the disease progression?
2. What EHR-derived data types and risk factors (eg, physiological, genetics, demographics) have been used as features for predictive modeling?
3. What are the research foci of the reviewed articles that use ML methods on EHR-derived data for modeling and predicting the progression of AD dementia?

### Search strategy

The aim of this SLR is to review works that meet the following criteria: (1) focus on modeling and predicting the onset or progression of AD dementia; (2) use ML techniques; and (3) use clinical markers of patients diagnosed with AD dementia. Similar to ref,[19] we created 3 keyword groups as follows, each relevant to different aspects of the scope of the review.

1. Keywords related to disease: AD, Alzheimer's, Alzheimer, Alzheimer's disease, dementia, Alzheimer's Disease, and Related Dementia.
2. Keywords related to ML methodology: ML, machine learning, AI, artificial intelligence, pattern recognition, computer-aided-diagnosis, CAD, classification, prediction, supervised learning, unsupervised learning, predictive modeling.
3. Keywords related to data and features: electronic health records, EHR, clinical data, clinical assessments, patient health data.

For each of the 3 keyword groups, we selected the words as per standard terminologies used for manuscript notation in the targeted literature databases.[19,21] The disease group consisted of different words related to AD and dementia. We observed that the terms "Mild Cognitive Impairment" and "MCI" often relate to brain diseases other than AD, so we excluded those to avoid false positives. For the ML methodology group, we focused on all possible variants related to ML/AI as well as general terms like prediction, classification, etc. The third group related to data and features consisted of keywords related to clinical EHR data. Since the focus of the review was on the use of clinical EHR-derived data with/or without imaging features, keywords related to imaging like "neuroimaging," "MRI," "PET," "CT," etc., were not part of any inclusion or exclusion criteria.

For our SLR, we searched the following bibliographic databases: (1) PubMed; (2) Association for Computing Machinery Digital Library; (3) IEEE Explore Digital Library; (4) ScienceDirect; (5) Scopus; and (6) arXiv/BioarXiv. Works originally identified from arXiv/BioarXiv were subsequently verified to have been accepted in a peer-reviewed journal or conference. Using each search engine, we searched the titles, abstracts, and keyword sections of articles published in journals or conference proceedings between January 1, 2010, and May 31, 2020. In order to limit our search to the scope of the review, our search string in each of the online databases was a triplet with 1 keyword from each of the 3 groups. All possible string combinations were created by taking 1 term from each of the 3 keyword groups joined by an "AND." The set of above formed triplets were then used as queries for the search.

### Exclusion criteria

The entire procedure of article searching and inclusion/exclusion criteria was performed according to the PRISMA (Preferred Reporting Items for Systematic Reviews and Meta-Analyses) guidelines.[22] The inclusion/exclusion criteria were reviewed by a board-certified neurologist and dementia specialist. All articles were selected and screened for eligibility by a doctoral student. In the first phase of screening, all duplicate articles collected from the different source libraries were removed. The next phase of screening discarded all papers that were clearly not relevant to the review, including studies where the abstract and the title did not contain any of the keywords related to "Alzheimer disease" or "Machine Learning." Following the approach adopted by the authors in ref,[21] the articles that passed the screening phase were assessed for eligibility by reviewing the full texts of the remaining articles to exclude studies which met one or more of the exclusion criteria (Table 1).

### Study risk of bias assessment

To mitigate the risk of bias during the search process and inclusion/exclusion of the articles, a series of checks were implemented during the article selection process. All articles were selected and screened for eligibility by a doctoral student. Two additional authors validated the final set of papers and review analysis. The final selected articles and inclusion/exclusion criteria were also reviewed by a board-certified neurologist and dementia specialist for relevance to our main research question.

### Summary statistics

Replicating the method followed in ref,[21] we calculated the following summary statistics from the final set of included articles: (1) source and publication year of article; (2) research focus of the article; (3) modality and accessibility of dataset; (4) size of cohort and type of features/risk factors; and (5) type of ML model for predictive modeling.

## RESULTS

Figure 1 shows the PRISMA flowchart, which depicts the selection process by which we arrived at the final set of included studies. We identified a total of 1331 studies from the different bibliographic databases. Since many papers were included in multiple databases, the first exclusion step removed 405 duplicate articles. From the remaining 926 articles, we removed 345 studies that were out of the scope of our review. This includes articles where the title or abstract did not contain any of the keywords related to "Alzheimer disease" or "Machine Learning." We reviewed 581 full-text articles for eligibility based on the exclusion criteria described in Table 1. After filtering on one or more of the exclusion criteria in Table 1, 64 articles remained for review.

Figure 2 shows the distribution of the reviewed articles by their publication year. The value shown for the year 2020 only includes data until May, when we performed the search. As shown by the plot, the count of published papers relevant to our scope has increased over the past 5 years, demonstrating that interest in using ML for analyzing AD with clinical data is on an upward trajectory.

### Data characteristics

To understand the nature of the data used by researchers in their articles, we documented the accessibility of the dataset, the number of included human subjects, and the incorporated clinical features. For each of the articles, we checked if the authors provided directions on how to access the dataset used in their experiments. We observed 2 main categories of datasets from our analysis: (1) deidentified datasets that are publicly available for download and (2) restricted datasets from sources like institutional clinical datasets that are not available for public use.

For the first category, we found that the Alzheimer's Disease Neuroimaging Initiative (ADNI) dataset was the most widely used, with 64% (41/64) of articles using ADNI for longitudinal AD data. ADNI enrolls participants between the ages of 55 and 90 at sites in the United States and Canada. After obtaining informed consent, participants undergo a series of neuropsychological and clinical assessments, genetic testing, and imaging (MRI and PET) at multiple

**Table 1.** Exclusion criteria for research articles

| | Exclusion criteria | Reasons for exclusion |
| --- | --- | --- |
| 1. | Only neuroimaging features were considered for predictive modeling | Scope of the review was inclusion of clinical EHR-derived data with/or without imaging features |
| 2. | ML methods were not used for clinical predictive modeling related to AD/dementia | We excluded articles which performed cohort summarization and hypothesis testing using statistical methods like logistic regression odds ratio, Chi-square distribution, Kruskal–Wallis test, etc. |
| 3 | Focus on a disease other than AD/dementia | AD/dementia is not the focus of the main analyses |
| 4. | AD/dementia is used only as an example of a neurodegenerative disease | AD/dementia is not the focus of the main analyses |
| 5. | Not peer-reviewed conference proceedings, journal, or preprints | Outside the scope of our review |
| 6. | Multiple publications from the same research group with similar final outcomes. In such cases, only the most recent studies were considered | Considered to be duplicate articles |
| 7. | Review articles | Review articles did not focus on a specific research goal |

*Abbreviations*: AD: Alzheimer disease; EHR: electronic health record; ML: machine learning.

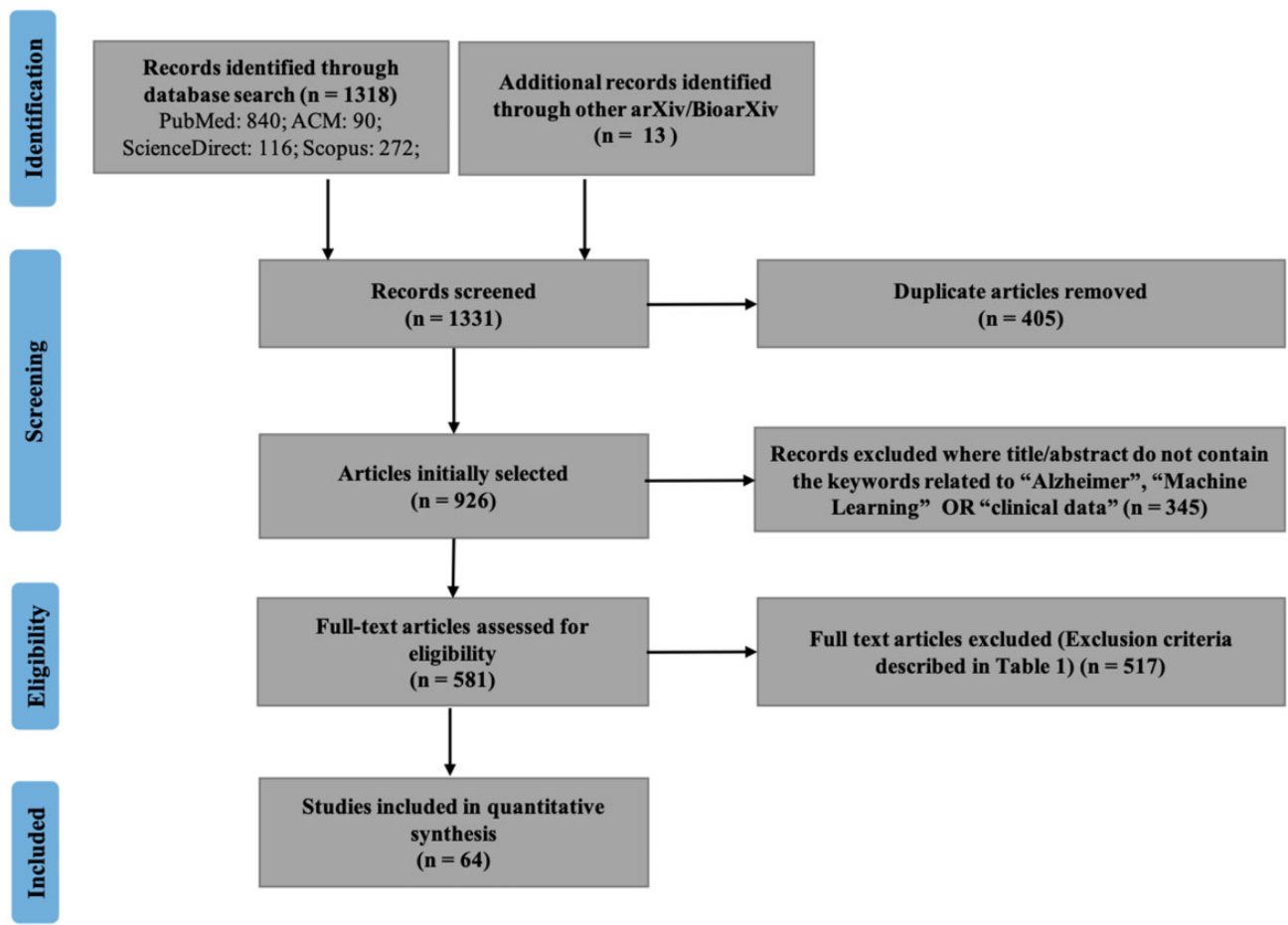

**Figure 1.** PRISMA flow diagram. *Abbreviation*: ACM: Association for Computing Machinery.

timepoints over subsequent years.[23] Other data sources included the National Alzheimer's Coordinating Center;[24–26] Australian Imaging, Biomarker & Lifestyle Flagship Study of Ageing;[27] Framingham Heart Study;[28,29] and Coalition Against Major Databases.[30] These datasets are all publicly available for download with some requiring a license from their respective websites.

For the second category, 16 out of 64 papers used their own customized clinical datasets. We deemed such datasets "restricted data" when there were no references or external links through which the data could be accessed. Examples of restricted datasets analyzed in these papers included a dataset subsampled from the National Health Insurance Service—national sample cohort of 1 million people representative of the South Korean population within the Korean National Health Insurance Service database[31] and a dataset collected from people who received a screening test in the dementia center in Gangbuk-Gu, Seoul, from 2008 to 2013.[32]

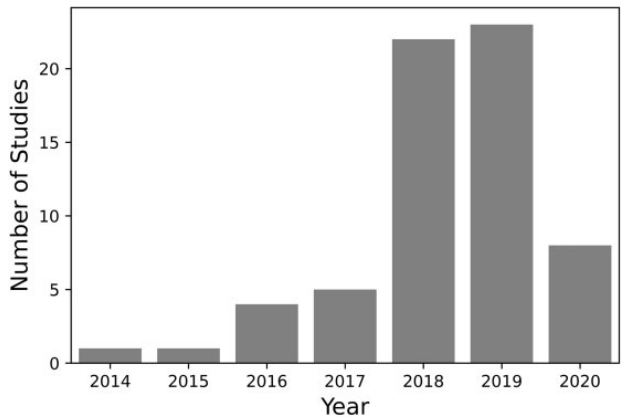

**Figure 2.** Studies published per year which use machine learning on clinical data for prognostic estimates of Alzheimer's Disease. For 2020, studies dated till May 31st were considered for the review.

The performance of an ML framework depends significantly on the size of the training cohort. From each of the reviewed articles, we determined the number of patients whose health data were used for analysis of AD dementia. Based upon the observed cohort sizes in all the included studies, we divided the cohort sizes into 4 categories: (1) 0–1000 patients; (2) 1001–10 000; (3) 10 001–100 000; and (4) >100 000. Figure 3 shows the number of studies corresponding to each cohort size category. Three articles[33–35] did not report the cohort sizes; so, we excluded them from our analysis of Figure 3. Most of the studies had cohort sizes of 0–1000 patients ($n = 34$), followed by 1001–10 000 patients ($n = 15$), 10 001–100 000 patients ($n = 5$), and finally >100 000 patients ($n = 7$).

### AD dementia features and biomarkers

Data of patients analyzed for risk of AD dementia consist of clinical variables like laboratory results, vital signs, neurobehavioral status exam scores, demographic information, and comorbidities, along with neuroimaging scans and CSF biomarkers. ML models try to learn the relationships amongst a set of clinical variables and determine if and how these variables contribute to the model predicting the development of AD dementia. Although the scope of this review focused on articles using nonimaging clinical data, some of these studies used multimodal datasets with nonimaging clinical and imaging features. To determine the prevalence of the use of nonimaging clinical features as potential AD risk factors, we classified the included studies into the following 2 categories: (1) Clinical only— only nonimaging clinical variables[36,37] and (2) Clinical + Imaging— imaging and nonimaging clinical variables were integrated to form the complete set of features.[38–41] We grouped the features into the following categories: neuroimaging features,[35,42,43] cognitive assessments,[44–47] genetic factors,[48–51] laboratory test values,[52–54] patient demographics,[55–57] and clinical notes.

Table 2 summarizes the feature categories and measures/factors used while applying the ML framework along with the count of articles using that particular variable. In our cohort of studies that used data from neuroimaging techniques as features for predictive modeling, MRI was the most widely used.[35,42,43] As shown in Table 2, cognitive assessments (48 studies) and demographics (47 studies) were the 2 most common features used by researchers for analysis of AD dementia. Only 4 articles considered clinical notes, primarily patient medical history and diagnosis details documented by clinicians.[58–61] Thirty studies (47%) were categorized as Clinical only and 34 (53%) as Clinical + Imaging. Figure 4 shows the relationship between the nature of data access restrictions (publicly available or restricted) and the category of AD dementia features (Clinical only or Clinical + Imaging). As illustrated by the figure, 94% (32/34) of the Clinical + Imaging data were extracted from publicly available datasets. For the Clinical only studies, 57% (17/30) of the Clinical data originated from datasets which are not publicly accessible.

### Application of ML methods

ML tools can model complex relationships between different clinical variables that are often beyond human capabilities. The output of trained ML models when applied on previously unseen healthcare clinical data yield inferences that can augment clinical decision-making. With the advancement of computational resources, researchers have progressed from simple ML algorithms like regression to complex deep learning models. We examined the different categories of ML techniques used in the reviewed articles based on the model type and the type of learning algorithm. We grouped the ML methods based on model type into the following categories: regression,[52,62,63] SVM,[64–66] decision tree,[67–69] Bayesian networks,[70–72] neural networks,[33,73,74] and natural language processing (NLP). The neural networks category includes both classifiers such as multilayer perceptron and deep learning models such as convolutional neural networks and autoencoders.[75–77] Based on the type of learning algorithm, ML models can either be supervised, unsupervised, or semi-supervised. In a supervised learning model, the algorithm is trained on a dataset annotated with gold standard labels. Unsupervised learning models, on the other hand extract features and patterns from unlabeled data and cluster the data points into distinct classes. Semi-supervised learning is a hybrid of the above 2 methods and combines a small amount of labeled data with a large volume of unlabeled data during training. Regression, SVM, decision trees, Bayesian networks, and neural networks all fall under the supervised category. k-Means algorithm is an example of unsupervised learning.[60]

Table 3 summarizes the types of ML models based on model type along with their different variants. The most widely used techniques were decision trees (50%), neural networks (44%), regression (34%), SVM (34%), and Bayesian networks (20%). NLP was

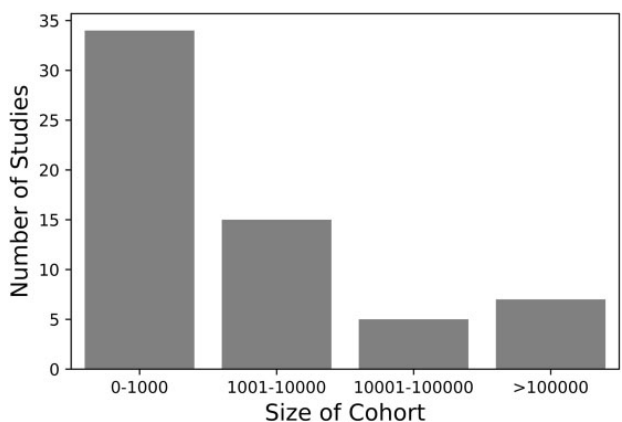

**Figure 3.** Size of cohort used in the reviewed studies.

**Table 2.** Features related to AD dementia identified by articles

| Feature categories | Measures/factors | Number of articles (%) |
| --- | --- | --- |
| Neuroimaging | MRI (structural, functional, unspecified), PET (FDG, amyloid) | 35 (54%) |
| Cognitive assessments | MMSE, ADAS-Cog, others (CDR, FAQ) | 48 (75%) |
| Genetic | *APOE ε*4, family history | 24 (38%) |
| Laboratory | CSF, vitals, medications, medical history, other laboratory tests | 32 (50%) |
| Demographics | Age, gender, education, race | 47 (72%) |
| Clinical notes | Discharge summary | 4 (6%) |

*Abbreviations:* AD: Alzheimer disease; ADAS-Cog: Alzheimer's Disease Assessment Scale-cognitive subscale; *APOE ε*4: apolipoprotein epsilon 4 allele; CDR: clinical dementia rating; CSF: cerebrospinal fluid; FAQ: Functional Activities Questionnaire; FDG, •••; MMSE: Mini-Mental State Exam; MRI: magnetic resonance imaging; PET: position emission tomography.

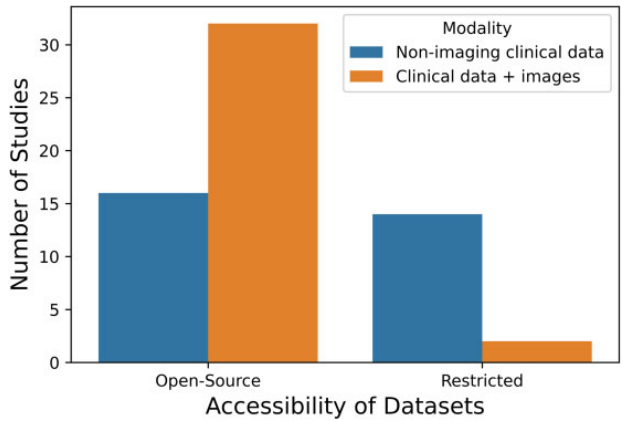

**Figure 4.** Relationship between the modality and accessibility of the datasets used in the included studies.

used only in studies which included patient clinical notes as one of the features; these comprised 6% of all studies.[58–61]

### Research foci of the reviewed articles

Identifying the research foci or the end goals of the reviewed articles can provide insights into how applying ML techniques on clinical EHR data can lead to effective clinical decision-making. More than half of the reviewed articles (55%, 35/64) investigated the progression of AD dementia to determine if an individual had stable or progressive AD. They aimed to predict the development of AD dementia in individuals who were initially cognitively normal or had only mild cognitive impairment.[34,78,79] Eleven/sixty-four (17%) studies used data comprised of longitudinal trajectories of clinical variables from patients showing mild symptoms of dementia to train predictive models for personalized forecasting of disease progression.[80–82] Eighteen/sixty-four (28%) studies in our review aimed to satisfy both the above objectives. For example,[83,84] presented a computational ML-based framework for modeling symptom trajectories using cognitive assessment scores at multiple time points and subsequently predicting those trajectory classes using multimodal data comprising both clinical and imaging variables.

### Reproducibility

Reproducibility, a fundamental requirement of the scientific process, is related to the idea that a scientific experiment should be able to be reproduced to validate its results.[85] The reproducibility of a scientific study is often assessed by the extent to which it follows the FAIR (Findable, Accessible, Interoperable and Reusable) principles.[86] For our review, we analyzed if the authors have provided adequate details about the dataset used and the implementation to check if the FAIR principles were followed. Only 7 (11%) of the articles reported their implementation code[27,37] and 48 (75%) studies used publicly available datasets.[36,38,41]

## DISCUSSION

AD is the most common cause of dementia, which is a syndrome of impaired memory and/or thinking that interferes with activities of daily life. We performed an SLR of studies applying ML to clinical data to identify factors that predict risk for progression of AD dementia. There has been an exponential increase in such paper applying ML for AD over the past 3 years. We reviewed the selected articles according to key components of ML analysis such as data characteristics, computational algorithms, and research focus; in the process we identified gaps in the existing literature and potential opportunities for future research. Our review suggests that most of the articles focus on predicting the progression of the disease based on standardized publicly available multimodal datasets which include both neuroimaging and some nonimaging clinical data. Most commonly used nonimaging clinical features for predictive modeling include neurobehavioral status exam scores, patient demographics, neuroimaging data, and laboratory test values.

Clinical databases which are collected for specific research purposes (eg, data in clinical registries) or cleaned and curated to enhance data reusability (eg, MIMIC database[87]) are often relatively well-structured, standardized, and clean, even though they may still have a few missing values and outliers. Hence, many researchers focus on utilizing these relatively clean datasets for their experiments and methodological innovations. However, as we noted previously, clinical data from local sources like institutional EHRs, which are primarily used to track patient care but can also be used secondarily for clinical research and automated disease surveillance, have great potential for use in modeling AD dementia progression.[88] Data from such raw EHR data sources often have data quality issues and require significant effort for data preprocessing and feature engineering. However, they are a rich source of historical clinical data containing patient-level elements which can be effectively leveraged using ML-based computational techniques for longitudinal analyses of their preclinical phase to identify prognostic clinical phenotypes, thus representing an opportunity to employ precision medicine paradigms in disease states where the current evidence-base precludes such an approach.

The basic criteria for selecting articles for our review was inclusion of clinical data excluding imaging with/or without other features and/or data types. We observed that a significant portion of articles employed neuroimaging features from structural and functional MRI

**Table 3:** Specific computational and machine learning methods utilized

| Computational methods | Specific models | Number of articles (%) |
| --- | --- | --- |
| Regression | Linear regression | 22 (34%) |
| | Logistic regression | |
| | Lasso regression | |
| | Ridge regression | |
| | Support vector regression | |
| SVM | SVM with linear kernel | 22 (34%) |
| | SVM with RBF kernel | |
| | SVM with polynomial kernel | |
| | Support vector regression | |
| Decision trees | Decision trees | 32 (50%) |
| | Random forest | |
| | Adaboost | |
| | GBM | |
| Bayesian networks | Naïve Bayes model | 13 (20%) |
| | Bayesian belief networks | |
| | GMM | |
| Neural networks | Multilayer perceptron | 28 (44%) |
| | CNN-based models | |
| | RNN-based models | |
| | Autoencoder | |
| | RBM | |
| | Graph neural networks | |
| NLP | Text mining | 4 (6%) |
| Others | KNN | 7 (11%) |
| | k-Means | |

*Abbreviations*: CNN: convolutional neural network; GBM: gradient boosting models; GMM: Gaussian mixture model; KNN: K-nearest neighbor; NLP: natural language processing; RBF: radial basis function; RBM: restricted Boltzmann machines; RNN: recurrent neural network; SVM: support vector machines.

as well as fluorodeoxyglucose (FDG) and amyloid positron emission tomography (PET) for predictive modeling; this indicates that most researchers use clinical features as part of multidimensional datasets containing both clinical and imaging features. As evident from the relationship between the modality and accessibility of the datasets, multimodal features are mostly derived from the category of publicly accessible, standardized, and well-curated datasets.

Identifying individuals with early AD brain pathological changes could enable therapeutic interventions to delay the disease progression over time and can be helpful for tailoring disease management and planning future care. Multiple failed drug trials for AD dementia show that in the later stages of the disease course, when the patient already has significant neuronal degeneration, treatment is unlikely to be helpful.[89] Hence, many drug trials are now enrolling patients with either preclinical AD (cognitively normal individuals with AD-related brain pathology) or very early AD dementia.[90,91] Most studies reported that their proposed methodology can identify individuals at risk for progression to AD dementia approximately 24–48 months before the diagnosis of AD dementia.

Nearly all of the reviewed articles used supervised learning in the proposed models. Unsupervised or semi-supervised learning can also be an effective tool for handling multidimensional longitudinal patient data where clinical outcomes are not known *a priori*. Unsupervised clustering algorithms can be helpful for identifying novel subphenotypes with distinct disease trajectories and the associations between them.[89,92]

EHR-derived data for patients who are screened for risk of AD dementia not only include structured data in the form of labs, medications, and procedures, but also clinical notes, which are textual descriptions of physician–patient encounters and records of their follow-up visits.[93] We observed from the summary statistics that information from clinical notes are not often included for developing the predictive modeling pipelines. However, these notes often consist of additional clinical information that are usually unavailable in the structured data sources, offering a rich source of information for clinical decision-making. Most of the notes are free-text narratives lacking a standardized structure and they cannot be processed by conventional ML algorithms like SVM, decision trees, regression, etc. NLP is a field of computational techniques that offers a viable solution for effectively processing clinical notes. In recent years, deep learning-based NLP models like recurrent neural networks and long short-term memory networks have been shown to outperform the conventional word-embedding-based NLP techniques for extracting relevant information from clinical notes.[94]

Data sharing and reproducibility of results can also enhance the impact, adaptation, and generalizability of research. Ideally, measures such as use of standardized publicly available EHR-derived datasets and specification of implementation details can help ensure reproducibility of the published methods and results. However, siloed data between different academic and corporate institutions and inconsistent data formats often make data sharing difficult and therefore remains an open area of research and innovation.[95] A solution to this problem is developing a culture of data sharing among different institutions, potentially utilizing common data models like the OHDSI (Observational Health Data Sciences and Informatics) data sharing initiative. OHDSI produces tools like the OMOP (Observational Medical Outcomes Partnership) Common Data Model, which transforms data within disparate observational databases into a common representation (terminologies, vocabularies, and coding schemes) so that they can be analyzed using standardized analytics tools.[96]

Despite these limitations, there have been significant advancements in the application of ML on EHR-derived data for predicting

AD dementia progression over the last 2–3 years. Advancements in technology and computational tools provide an opportunity for researchers to develop deep learning-based computational hypotheses that can inform clinical decision-making. Deep learning and other analytic approaches in ML can define clinical patterns and generate insights beyond human capabilities. This not only reduces the burden on clinicians in making their diagnoses but leads to improved quality, safety, and outcomes of care planning and delivery.[18]

### Limitations of SLR

Many relevant research works might be published in only conferences and workshops proceedings and not indexed in bibliographic databases. Similarly, some studies from arXiv/BioarXiv were not included since they were not yet peer-reviewed. Thus, identification of only peer-reviewed studies from bibliographic databases might lead to selection bias during the initial inclusion stage. This can potentially impact the results presented as it may not truly represent the growing interest in the domain of using ML models for AD prediction using clinical data.

## CONCLUSION

We performed an SLR of studies using ML on clinical EHR data for modeling and prediction of AD dementia progression. We summarized different aspects of the articles including data source and modality, features, methods, and research focus of the studies. The summarized results suggest that most state-of-the-art research on AD has focused on predicting the progression of the disease based on standardized publicly available multimodal datasets which include both neuroimaging and some nonimaging clinical data. The nonimaging clinical data used most commonly for predictive modeling include neurobehavioral status exam scores, patient demographics, neuroimaging data, and laboratory test values. Almost all the reviewed articles utilized supervised learning with common ML models such as neural networks, decision trees, SVM, and regression. ML and other analytic approaches in AI can generate helpful insights about complex clinical patterns that assists clinicians in their decision-making and leads to improved quality, safety, and outcomes of healthcare planning.

## FUNDING

The preparation of this report was supported by the Centene Corporation contract (P19-00559) for the Washington University-Centene ARCH Personalized Medicine Initiative. SS is supported by K23AG053426.

## AUTHOR CONTRIBUTIONS

SK and AG conceived and designed the study. SK did the data collection (literature review), analysis, and interpretation, drafted and revised the manuscript, and prepared the graphical illustrations. AG and IO participated in the literature review interpretation, drafted, and revised the manuscript, and approved the final version for submission. PROP, AML, and SS reviewed and revised the manuscript, and approved the final version for submission.

## SUPPLEMENTARY MATERIAL

Supplementary material is available at *Journal of the American Medical Informatics Association* online.

## CONFLICT OF INTEREST STATEMENT

None declared.

## DATA AVAILABILITY

The data underlying this article are available in the article and in its online supplementary material.